\documentclass[prd,aps,twocolumn,floats,floatfix,amsmath,nofootinbib]{revtex4-1}
\usepackage{graphicx,array,dcolumn}
\usepackage{calc,tabularx, epsfig,mathrsfs}
\usepackage{hyperref}

\usepackage{amsmath,verbatim,enumerate}
\usepackage{amssymb}
\usepackage{multirow}
\usepackage{xspace}
\usepackage{slashed}
\allowdisplaybreaks[1]
\newlength{\figurewidth}%
\newcommand{\beq}{\begin{equation}}
\newcommand{\eeq}{\end{equation}}
\newcommand{\bea}{\begin{eqnarray}}
\newcommand{\eea}{\end{eqnarray}}
\newcommand{\ba}{\begin{array}}
\newcommand{\ea}{\end{array}}

\newcommand{\mn}{{\mu\nu}}

\newcommand{\pt}{\partial}

\newcommand{\al}{\alpha}

\newcommand{\ta}{\theta}

\newcommand{\Lam}{\Lambda}

\newcommand{\OM}{\Omega}

\newcommand{\sg}{\sigma}

\makeatother
\begin{document}
%
\title{Cosmic evolution in novel-Gauss Bonnet Gravity}
\setlength{\figurewidth}{\columnwidth}
%
\author{Gaurav Narain$\,{}^a$}
\email{gaunarain@gmail.com}
\author{Hai-Qing Zhang$\,{}^{a,b}$}
\email{hqzhang@buaa.edu.cn}

\affiliation{${}^a$ Center for Gravitational Physics, Department of Space Science, 
Beihang University, Beijing 100191, China. \\
${}^b$ International Research Institute for Multidisciplinary Science, Beihang University, 
Beijing 100191, China}

%
\begin{abstract}
In this short paper we investigate any non-trivial effect the novel Gauss-Bonnet gravity 
may give rise in the cosmic evolution of the Universe in four spacetime dimensions. 
We start by considering a generic Friedmann-Lema\^itre-Robertson-Walker (FLRW) 
metric respecting homogeneity and isotropicity
in arbitrary space-time dimension $D$. The metric depends on two functions: scale factor and lapse.
Plugging this metric in novel Einstein-Gauss-Bonnet (EGB)
gravity action, doing an integration by parts and then take the limit 
of $D\to4$ give us a dynamical action 
in four spacetime dimensions for scale-factor and lapse. The peculiar rescaling 
of Gauss-Bonnet coupling by factor of $D-4$ results in a non-trivial contribution 
in the action of the theory. In this paper we study this action. 
We investigate the dynamics of scale-factor and behavior of lapse in an empty Universe
(no matter). Due to complexity of the problem we study the theory to first order 
in Gauss-Bonnet coupling and solve system of equation to the first order.
We compute the first order correction to the on-shell action of the empty Universe
and find that its sign is opposite of the leading order part. We discuss it 
consequences.
\end{abstract}
\maketitle

\section{Introduction}
\label{intro}

General relativity although is a very good theory of gravity which works over a large 
range of energy, but it seems to be lacking a complete description of gravity.
It is expected to get modified at short distances and/or in deep infrared. 
At short distances such modification are motivated as the quantum 
theory of GR has problems: theory is non-renormalizable and lacks 
the ability to predict \cite{tHooft1974,Deser19741,Deser19742,Deser19743,
Goroff1985,Goroff1985a,vandeVen1991}. 
In deep infrared the observations supporting 
dark-matter and dark energy \cite{Riess:1998cb,Perlmutter:1998np,ArmendarizPicon:1999rj,ArmendarizPicon:2000dh} 
are key motivations for modification of gravity 
at long distances \cite{Maggiore:2014sia}.

It is generally seen that any kind of modification introduced to 
Einstein-Hilbert gravity leads to some undesirable problems. 
For example at high-energies to address issues of renormalizability, 
a higher-derivative modification although resolves renormalizability 
\cite{Stelle:1976gc,Salam:1978fd,Julve:1978xn} but introduces 
a nasty problem of non-unitarity. Some efforts have been made 
to tackle it \cite{Narain:2011gs,Narain:2012nf,Narain:2017tvp,Narain:2016sgk}, 
and the direction has seen 
a recent uprising where the interest has been rekindled 
following works in asymptotic safety approach 
\cite{Codello:2006in,Niedermaier:2009zz,Benedetti:2009rx}
and `\textit{Agravity}' \cite{Salvio:2014soa}. Issues of unitarity 
arises in such higher-derivative gravity theories 
as the equation of motion contains more than two time-derivatives of metric. 
Lovelock gravity \cite{Lovelock:1971yv,Lovelock:1972vz,Lanczos:1938sf} 
are a special class of higher-derivative gravity 
whose equation of motion remains second order in time, and 
in fact Einstein gravity can be extended by a whole tower of 
higher-curvature terms where the equation of motion still remains 
second-order in time.

In four spacetime dimension the Lovelock gravity also known 
as Gauss-Bonnet gravity is topological and doesn't 
contribute in the dynamical evolution of metric. However, they 
play a central role in the path-integral quantization of gravity where 
it is used to classify topologies. Recently, it has been noticed in 
\cite{Glavan:2019inb} that the Gauss-Bonnet gravity term in the action 
can contribute non-trivially if its coefficient in action has be 
rescaled by $(D-4)$, where $D$ is spacetime dimensionality. This motivation for rescaling arises 
following works \cite{Mardones:1990qc,Torii:2008ru} where it was 
observed that Gauss-Bonnet contribution to equation of motion is proportional 
to $(D-4)$. Such rescaling introduces non-trivial features 
coming from Gauss-Bonnet in four spacetime dimensions. 
This has generated tremendous interest in novel Gauss-Bonnet gravity.

The novel Gauss-Bonnet gravity \cite{Glavan:2019inb} action is following
\bea
\label{eq:act}
S &=& \frac{1}{16\pi G} \int {\rm d}^Dx \sqrt{-g}
\biggl[
-2\Lam + R + \frac{\al}{D-4} 
\biggl( R_{\mu\nu\rho\sg} R^{\mu\nu\rho\sg} 
\notag \\
&&- 4 R_\mn R^\mn + R^2 \biggr)
\biggr] \, , 
\eea
where $G$ is the Newton's gravitational constant, 
$\Lam$ is cosmological constant term and 
$\al$ is the Gauss-Bonnet coupling. The Gauss-Bonnet coefficient 
is defined with a $(D-4)$ factor in denominator. 
$G$ has mass dimension $M^{2-D}$, $\Lam$ has mass 
dimension $M^2$, while $\al$ has mass dimension $M^{-2}$. 

By now the action in eq. (\ref {eq:act}) has been explored in the context of
spherical black holes \cite{Kumar:2020uyz,Fernandes:2020rpa,Kumar:2020owy}, 
star-like solutions \cite{Doneva:2020ped}, radiating solutions \cite{Ghosh:2020vpc}, 
collapsing solutions \cite{Casalino:2020kbt}, even extensions to more higher-curvature 
Lovelock gravity theories \cite{Casalino:2020kbt,Konoplya:2020qqh}. 
There are already a numerous investigation on the 
thermodynamic behavior \cite{Zhang:2020qam,Konoplya:2020cbv,Wei:2020poh} of these objects
quasi-normal modes \cite{Aragon:2020qdc}, strong cosmic-censorship \cite{Mishra:2020gce},
bending of light \cite{Heydari-Fard:2020sib}, lower-dimensional solutions 
\cite{Nojiri:2020tph,Hennigar:2020fkv}.
Using the recent observations of blackhole shadows constraints on its 
coupling parameter has also been investigated 
\cite{Wei:2020ght,Roy:2020dyy,Guo:2020zmf,Zeng:2020dco}.

It is noticed in \cite{Lu:2020iav,Mahapatra:2020rds,Kobayashi:2020wqy,Hennigar:2020lsl} that 
under an integration by parts the
novel Gauss-Bonnet gravity action gets rid of $(D-4)$ factors, thereby leading to 
a well-defined $D\to4$ limit at the level of action. Their strategy is inspired 
by Kaluza-Klein compactification where they decompose the metric in two 
parts ${\cal M}_D = {\cal M}_4 \bigotimes {\cal M}_{D-4}$: 
four-dimensional manifold ${\cal M}_4$ and extra-dimension piece ${\cal M}_{D-4}$. 
They write their 
action under this decomposition and smoothly remove the extra-dimension 
piece by taking limit. One then gets Horndeski  type gravity. A similar 
study was done in \cite{Kobayashi:2020wqy,1794944} using ADM decomposition. There they noticed 
that for a well-defined limit and a consistent theory in four dimensions
one either break (a part of) the diffeomorphism invariance or have an extra degree of freedom, 
in agreement with the Lovelock theorem \cite{1794944}.

Inspired by the above work we decided to investigate the cosmological 
settings in the novel-Gauss-Bonnet gravity.
We start by considering the 
most general metric respecting spatial homogeneity and isotropicity in $D$-spacetime 
dimensions. This is a generalisation of FLRW metric in $D$-dimensions 
consisting of two unknown time-dependent functions: lapse 
and scale-factor. This is different from the KK decomposition
that has been studied in \cite{Lu:2020iav,Hennigar:2020lsl,1794944}.
Plugging the generalised FLRW metric in action of theory and performing 
an integration by parts results in action where $D\to4$ limit 
can be smoothly taken without encountering divergences. 
This leaves us with a well-defined action for scale factor
and lapse, which gets non-trivial corrections from 
Gauss-Bonnet gravity part. In this paper we study this 
and investigate the dynamics of scale-factor and lapse 
in empty matter free Universe. 

The paper is organized as follows: in section 
\ref{flrw} we write the generic FLRW metric and compute its action, 
in section \ref{k0} we study the action for flat Universe and study the 
equation of motion perturbatively. We conclude with a 
discussion in section \ref{conc}.

\section{FLRW metric}
\label{flrw}

We start by considering generalization of FLRW metric in arbitrary spacetime 
dimension whose dimensionality is $D$. We write the metric in polar
co-ordinates $\{t_p, r, \ta, \cdots \}$
\beq
\label{eq:frwmet}
{\rm d}s^2 = - N_p^2(t_p) {\rm d} t_p^2 
+ a^2(t_p) \left[
\frac{{\rm d}r^2}{1-kr^2} + r^2 {\rm d} \OM_{D-2}^2
\right] \, ,
\eeq
where $N_p(t_p)$ is lapse function, $a(t_p)$ is scale-factor, 
$k=(0, \pm 1)$ is the curvature, and ${\rm d}\OM_{D-2}$ is the 
metric corresponding to unit sphere in $D-2$ spatial dimensions. 
This metric is conformally related to flat metric and hence
its Weyl-tensor $C_{\mu\nu\rho\sg} =0$. The nonzero entries 
of Riemann tensor 
\cite{Deruelle:1989fj,Tangherlini:1963bw,Tangherlini:1986bw,Cognola:2013fva} are 
\bea
\label{eq:riemann}
R_{0i0j} &=& - \left(\frac{\ddot{a}}{a} - \frac{\dot{a} \dot{N_p}}{a N_p} \right) g_{ij} \, , 
\notag \\
R_{ijkl} &=& \left(\frac{k}{a^2} + \frac{\dot{a}^2}{N_p^2 a^2} \right)
\left(g_{ik} g_{jl} - g_{il} g_{jk} \right) \, ,
\eea
where $g_{ij}$ is the spatial part of the FLRW metric
and $(\dot{})$ denotes derivative with respect to $t_p$.
The non-zero components of the Ricci-tensor are 
\bea
\label{eq:Ricci-ten}
R_{00} &=& - (D-1) \left(\frac{\ddot{a}}{a} - \frac{\dot{a} \dot{N_p}}{a N_p} \right)
\, , 
\notag \\
R_{ij} &=& \left[
\frac{(D-2) (k N_p^2 + \dot{a}^2)}{N_p^2 a^2}
+ \frac{\ddot{a} N_p - \dot{a} \dot{N_p}}{a N_p^3} 
\right] g_{ij} \, ,
\eea
while the Ricci-scalar is given by
\beq
\label{eq:Ricci0}
R = 2(D-1) \left[\frac{\ddot{a} N_p - \dot{a} \dot{N_p}}{a N_p^3} 
+ \frac{(D-2)(k N_p^2 + \dot{a}^2)}{2N_p^2 a^2} \right] \, .
\eeq
The good thing about FLRW metric is that it is Weyl-flat which implies 
that Riemann tensor can be expressed in terms of 
Ricci-tensor and Ricci scalar as follow 
\bea
\label{eq:Riem_exp}
R_{\mu\nu\rho\sg} &=& \frac{R_{\mu\rho} g_{\nu\sg} - R_{\mu\sg}g_{\nu\rho}
+ R_{\nu\sg} g_{\mu\rho} - R_{\nu\rho} g_{\mu\sg}}{D-2}
\notag \\
&&
- \frac{R (g_{\mu\rho} g_{\nu\sg} - g_{\mu\sg} g_{\nu\rho})}{(D-1)(D-2)} \,.
\eea
This identity is valid for all conformally flat metrics and allows one to express
\beq
\label{eq:Reim2_exp}
R_{\mu\nu\rho\sg} R^{\mu\nu\rho\sg}
= \frac{4}{D-2} R_\mn R^\mn - \frac{2 R^2}{(D-1)(D-2)} \, .
\eeq
By making use of this identity for conformally flat metrics in the 
Gauss-Bonnet action one can obtain a simplified action of the theory.
In such cases we have
\bea
\label{eq:actGB}
&&
\int {\rm d}^Dx \sqrt{-g} \left(
R_{\mu\nu\rho\sg} R^{\mu\nu\rho\sg} - 4 R_\mn R^\mn + R^2
\right)
\notag \\
&&
= \frac{D-3}{D-2} \int {\rm d}^Dx \sqrt{-g} \left(
- R_\mn R^\mn + \frac{D R^2}{D-1}
\right) \, .
\eea
On plugging the FRW metric of eq. (\ref{eq:frwmet}) in the action 
in eq. (\ref{eq:act}) one can get an action for $a(t_p)$ and 
$N_p(t_p)$. On doing integration by parts it is noticed that the 
resulting terms are independent of factors of $(D-4)$, which cancels off.
This resulting action in $D=4$ is given by
\bea
\label{eq:act_frw_int}
&&
S = \frac{\pi k^{-3/2}}{8 G} \int {\rm d}t_p \biggl[
(3 k -  \Lam a) N_p a - \frac{3a a^\prime}{N_p} 
\notag\\
&&
+ \frac{3\al}{a} \biggl\{
\frac{(kN_p^2 + a^{\prime2})^2}{N_p^3} + \frac{4k a^{\prime2}}{N_p}
+ \frac{4 a^{\prime4}}{N_p}
\biggr\}
\biggr] \, ,
\eea
where $({}^\prime)$ denotes derivative with respect to $t_p$.
Here one notices that the Gauss-Bonnet term gives a non-trivial contribution in 
$D=4$ which is possible due to particular style of defining the Gauss-Bonnet 
coupling parameter. With this action one can do further analysis. To make the theory 
more appealing one can rescale lapse and scale factor in following manner 
\beq
\label{eq:rescale}
N_p(t_p) {\rm d} t_p = \frac{N(t)}{a(t)} {\rm d} t \, ,
\hspace{5mm}
q(t) = a^2(t) \, .
\eeq
This transformation changes our original metric in eq. (\ref{eq:frwmet})
into following
\beq
\label{eq:frwmet_changed}
{\rm d}s^2 = - \frac{N^2}{q(t)} {\rm d} t^2 
+ q(t) \left[
\frac{{\rm d}r^2}{1-kr^2} + r^2 {\rm d} \OM_{D-2}^2
\right] \, ,
\eeq
and our action in eq. (\ref{eq:act_frw_int}) changes to following simple form.
\bea
\label{eq:Sact_frw_simp}
S &=& \frac{\pi k^{-3/2}}{16 G} \int {\rm d}t \biggl[
(6 k - 2\Lam q) N - \frac{3 \dot{q}^2}{2N}
\notag \\
&&
+ \frac{3\al}{8 N^3 q} (4k N^2 + \dot{q}^2)(4k N^2 + 5\dot{q}^2)
\biggl] \, ,
\eea
where $(\dot{})$ represent derivative with respect to $t$. It should be 
noticed that the action doesn't contains any derivative of $N$, which happens 
as we have performed integration by parts previously. This is an interesting 
higher-derivative action which only depends on $q$, $\dot{q}$ and $N$. 
This action further acquires simplicity when $k=0$. 

\section{Flat space $k=0$}
\label{k0}

The action acquires a simplified structure in flat space $(k=0)$, 
this is also motivated physically as our physical Universe is also 
observed to be spatially flat to a high accuracy. The action in this case 
is given by,
\beq
\label{eq:Sact_k0}
S_{k=0} = \frac{V_3}{16 \pi G} \int {\rm d}t \biggl[
- 2\Lam q N - \frac{3 \dot{q}^2}{2N}
+ \frac{15 \al \dot{q}^4}{8 N^3 q} \biggl] \, ,
\eeq
where $V_3$ is the volume corresponding to $3$-dimensional space.
As there is no derivative term corresponding to $N$ appearing in action 
so variation of action with respect to $N$ will result in a constraint. 
On the other hand, variation of action with respect to $q(t)$ will give 
evolution equation for $q(t)$. In the gauge $\dot{N}=0$ we have $N(t) = N_c$,
and the equation of motion for $q(t)$ is given by
\beq
\label{eq:eqmk0_1}
- 2 N_c \Lam + \frac{3 \ddot{q}}{N_c}
+ \frac{45\al}{8 N_c^3} \left(\frac{\dot{q}^4}{q^2} - \frac{4 \dot{q}^2 \ddot{q}}{q}
\right) =0 \, .
\eeq
This is a higher-derivative equation. The higher-derivative contribution is novel 
here which doesn't arise if the Gauss-Bonnet coupling wasn't rescaled by factor of 
$(D-4)$. In principle one has to solve for $q(t)$ from the above equation 
for the boundary conditions 
\beq
\label{eq:bound_q}
q(t=0) = b_0 \, , 
\hspace{5mm}
q(t=1) = b_1 \, .
\eeq
Then plug the $q(t)$-solution back in to the action in eq. (\ref{eq:Sact_k0})
where now we are in constant-$N$ gauge, integrate with respect to time
and obtain an action for the constant lapse $N_c$. One then look for saddle points solution for 
$N_c$ which are obtained by varying this action with respect to $N_c$. 
This will be the full saddle point solution of theory. 

In practice this is not always possible. In the current case 
the evolution equation for $q(t)$ is quite complicated and involve higher-derivates.
We therefore try to solve the system perturbatively. We start by 
expanding $q(t)$ in powers of $\al$.
\beq
\label{eq:qt_exp}
q(t) = q_0(t) + \al q_1(t) + \cdots \, ,
\eeq
where $q_0$ is zeroth-order solution while $q_1$ is the first order solution. 
%
\subsection{zeroth-order}
\label{0order}

At the zeroth order we have 
\beq
\label{eq:q0_evol}
\ddot{q} = \frac{2 N_c^2 \Lam}{3} \, . 
\eeq
It is a linear second order ODE which can be solved analytically exactly. 
The solution obeying the boundary condition stated in eq. (\ref{eq:bound_q})
is given by
\beq
\label{eq:q0_sol}
q_0(t) = \frac{\Lam N_c^2}{3} (t^2 -t) + b_0(1-t) + b_1 t \, .
\eeq
Plugging it back into action in eq. (\ref{eq:Sact_k0}) and integrating with respect to 
$t$ one gets zeroth-order part of action for $N_c$. This is given by
\beq
\label{eq:S0_act_N}
S_0 = \frac{V_3}{16\pi G} 
\left[-\frac{3(b_0-b_1)^2}{2N_c} 
- (b_0+b_1)N_c \Lam + \frac{N_c^3 \Lam^2}{18} \right]\, .
\eeq
Once we have the action for $N_c$, one can obtain saddle points of this 
by varying it with respect to $N_c$ and looking for extrema. Then we see 
that $\pt S_0/\pt N_c =0$ on solving gives $N_0$. 
Its solutions obey the equation 
\beq
\label{eq:Nc0_sad}
\frac{3(b_0-b_1)^2}{2N_0^2} 
- (b_0+b_1)\Lam + \frac{N_0^2 \Lam^2}{6} = 0 \, .
\eeq
This is a quadratic in $N_0^2$ and will consist of four solutions
which are given by
\beq
\label{eq:Nc0_soln}
\left(N_0\right)_{\pm, \pm}
= \pm \sqrt{\frac{3}{\Lam}}\left(\sqrt{b_1} \pm \sqrt{b_0} \right) \, .
\eeq
At the zeroth order we notice that there are four saddle point solutions. 
At this level we don't receive any correction from the Gauss-Bonnet term 
and they agree with the known saddles in the context of Lorentzian quantum cosmology
\cite{Feldbrugge:2017kzv,Feldbrugge:2017fcc}. Corresponding to each $\left(N_0\right)_{\pm\pm}$ 
we have corresponding $\left(q_0\right)_{\pm\pm}$. Each one of them lead to 
a different FLRW metric. Corresponding to each of them we have an on-Shell 
action, which is given by
\beq
\label{eq:S0act_onshell}
S_0^{\rm on-shell} = \mp \frac{V_3}{4 \pi G} \sqrt{\frac{\Lam}{3}}
\left(\sqrt{b_1^3} \pm \sqrt{b_0^3} \right) \, .
\eeq

\subsection{First order}
\label{first order}

At first order in $\al$ the equations becomes more involved. The evolution of 
$q(t)$ at first order is dictated by following equation 
\beq
\label{eq:q1_evo_eq}
\ddot{q}_1 = - \frac{15}{8 N_c^2} \left(
\frac{\dot{q}_0^4}{q_0^2} - \frac{4 \dot{q}_0^2 \ddot{q}_0}{q_0}
\right) \, , 
\eeq
where $q_0$ is the zeroth order solution to $q(t)$ obtained above. The boundary 
conditions for $q_1(t)$ can be obtained from eq. (\ref{eq:bound_q}) and 
those of $q_0$. This implies that 
\beq
\label{eq:boundQ1}
q_1(t=0) = q_1(t=1) = 0\, .
\eeq
The ODE for $q_1$ satisfying the these boundary conditions can be solved and its
solution is given by
\bea
\label{eq:q1sol}
&&
q_1(t) = \frac{5 N_c^2 \Lam^2 t(t-1)}{3} 
- \frac{5 U}{4 N_c^2} \biggl[
\left(b_0-b_1 + \frac{N_c^2\Lam}{3} \right)(t-1) 
\notag\\
&&
\times 
\tan^{-1} \left(\frac{3(b_0-b_1) + N_c^2\Lam}{U}\right)
+ \left(b_0 - b_1 - \frac{N_c^2\Lam}{3} \right)
\notag\\
&&
\times
t \tan^{-1} \left(
\frac{3(b_1-b_0) + N_c^2\Lam}{U}
\right)
+ \biggl(b_1 - b_0 
\notag\\
&&
+ \frac{N_c^2\Lam(2t-1)}{3} \biggr)
\tan^{-1} \left(\frac{3(b_1-b_0) + N_c^2 \Lam (1-2t)}{U} \right)
\biggr] \, ,
\eea
where 
\beq
\label{eq:Uform}
U = \sqrt{6(b_0+b_1)^2 N_c^2 \Lam - 9(b_0-b_1)^2 - N_c^4 \Lam^2} \, .
\eeq
Once we have obtained the first order correction to $q(t)$, we can plug 
it back in action in eq. (\ref{eq:Sact_k0}) and perform the $t$-integration. 
This results in a first order corrected action for $N_c$. 
\bea
\label{eq:S1act_N}
&&
S_1 = S_0 + \frac{V_3 \al}{16\pi G} 
\biggl[
\frac{5 (b_0 - b_1)^2 \Lam}{N_c} - \frac{5(b_0+b_1) N_c \Lam^2}{3}
\notag \\
&&
+ \frac{10N_c^3 \Lam^3}{27}
+ \frac{15 U^3 }{108 N_c^3} \biggl\{\tan^{-1} \left(\frac{3(b_0-b_1) + N_c^2\Lam}{U}\right)
\notag \\
&&
+ \tan^{-1} \left(\frac{3(b_1-b_0) + N_c^2\Lam}{U}\right) \biggr\}
+ \cdots 
\biggr] \, .
\eea
This first order corrected action can be varied with respect to $N_c$ again 
to obtain the first order correction to saddle points. To obtain this 
we substitute 
\beq
\label{eq:Nexp}
N_c = N_0 + \al N_1 + \cdots \, .
\eeq
Then $N_1$ can be obtained from  
\bea
\label{eq:N1sol_eq}
\left. \frac{\pt S_0}{\pt N_c} \right|_{N_c \to (N_0 + \al N_1)} = 0 \, .
\eea
From this equation we have 
\bea
\label{eq:N1_sol}
&&
N_1 = -\frac{5N_0 \Lam}{2} \, ,
\eea
where $N_0$ is given by eq. (\ref{eq:Nc0_sad}). Once the expression of 
$N_0$ given in eq. (\ref{eq:Nc0_sad}) is plugged back in above, we get
the full first order corrected lapse which is given by
\bea
\label{eq:N1_final}
&&
N_c
= s_1\sqrt{\frac{3}{\Lam}}
\left(1 - \frac{5 \al \Lam}{2} + \cdots \right) 
\left(\sqrt{b_1} + s_2 \sqrt{b_0} \right)  \, ,
\eea
where $s_{1,2} = \{\pm, \pm\}$. The first order corrected on-shell action is given by
\bea
\label{eq:S1_onshell}
&&
S_1^{\rm on-shell} = \left(1 - \frac{5\al \Lam}{6} + \cdots \right)
S_0^{\rm on-shell} \, ,
\eea
where $S_0^{\rm on-shell}$ is the zeroth order on-shell action 
given in eq. (\ref{eq:S0act_onshell}). 

\section{Discussion and conclusion}
\label{conc}

In this short paper we investigate the non-trivial effects that arise in novel-Gauss-Bonnet 
gravity, where the coefficient in front of the Gauss-Bonnet term in action has been 
appropriately rescaled by factor of $(D-4)$. We investigate cosmological spacetimes 
respecting homogeneity and isotropicicity. We consider a most general spacetime 
respecting such symmetry in $D$-spacetime dimensions. It consists of two functions:
scale factor $a(t_p)$ and lapse $N_p(t_p)$. Plugging this in the gravitational action, doing 
integration by parts and taking the limit $D \to 4$, we are able to obtain a dynamical 
action for the scale factor $a(t_p)$. We notice that an integration by parts (and discarding 
a surface term) allows us to obtain an action for the resulting theory  
where the $D-4$ terms cancels off completely. This allows us to perform a smooth 
$D\to 4$ limit without witnessing divergences. 
Similar observations were also made in \cite{Lu:2020iav,1794944} where 
the authors considered KK decomposition of metric. 
This interesting observation indicates that probably something 
deep is happening which requires more 
careful analysis. 

Once the action of theory is obtained one can study various aspects of it. 
We notice that a redefinition of scale factor and lapse lead us to a simplified 
theory for the new variables: $q(t)$ and $N(t)$. The resulting action is a function of 
$N$, $q$ and $\dot{q}$ only. The action doesn't contain any derivative term for the lapse,
indicating that its dynamical nature comes from the dynamics of $q(t)$ indirectly.
On varying the action with respect to $q(t)$ we obtain equation of motion. This is 
a non-linear higher-derivative ODE, and it might be hard to find an analytic closed form 
solution of same. We solve the system perturbatively, doing perturbation in the Gauss-Bonnet
coupling. We compute the first order correction to Einstein-Hilbert solution for
$q$ and $N$, for the case of empty Universe. 
Using these we also compute the first order correction to the on-shell action.
We notice that the sign of the first order correction to on-shell action is negative. This can have 
serious consequences in euclidean path-integral as such correction where $\al>0$ will 
lead to exponential growth. This perhaps maybe considered a shortcoming of this theory. 
Moreover, this is will also indicate that only one sign of coupling is allowed. 

Given that we have an action of theory in four spacetime dimensions it is then natural to 
investigate the quantum gravity path-integral and the corrections it achieves 
in the novel Gauss-Bonnet gravity. Full gravitational path-integral although quite 
complicated to deal with, still many nice features of it can be understood 
when degree of freedoms are reduced substantially. We will present this 
in our next publication \cite{nextpub}.

\bigskip
\centerline{\bf Acknowledgements} 
GN will like to thank Nirmalya Kajuri and Avinash Raju for useful discussion. 
GN is supported by ``Zhuoyue" Fellowship (ZYBH2018-03).
H. Q. Z. is supported by the National Natural Science Foundation of China (Grants
No. 11675140, No. 11705005, and No. 11875095).

\appendix



\begin{thebibliography}{99} 
%
%
\bibitem{tHooft1974} 
  G.~'t Hooft and M.~J.~G.~Veltman,
  ``One loop divergencies in the theory of gravitation,''
  Ann.\ Inst.\ H.\ Poincare Phys.\ Theor.\ A {\bf 20}, 69 (1974).
%
%
\bibitem{Deser19741} 
  S.~Deser, H.~S.~Tsao and P.~van Nieuwenhuizen,
  ``Nonrenormalizability of Einstein Yang-Mills Interactions at the One Loop Level,''
  Phys.\ Lett.\  {\bf 50B}, 491 (1974).
  doi:10.1016/0370-2693(74)90268-8
%
%
\bibitem{Deser19742} 
  S.~Deser and P.~van Nieuwenhuizen,
  ``One Loop Divergences of Quantized Einstein-Maxwell Fields,''
  Phys.\ Rev.\ D {\bf 10}, 401 (1974).
  doi:10.1103/PhysRevD.10.401
%
%
\bibitem{Deser19743} 
  S.~Deser and P.~van Nieuwenhuizen,
  ``Nonrenormalizability of the Quantized Dirac-Einstein System,''
  Phys.\ Rev.\ D {\bf 10}, 411 (1974).
  doi:10.1103/PhysRevD.10.411
%
%
\bibitem{Goroff1985} 
  M.~H.~Goroff and A.~Sagnotti,
  ``Quantum Gravity At Two Loops,''
  Phys.\ Lett.\  {\bf 160B}, 81 (1985).
  doi:10.1016/0370-2693(85)91470-4
%
%
\bibitem{Goroff1985a} 
  M.~H.~Goroff and A.~Sagnotti,
  ``The Ultraviolet Behavior of Einstein Gravity,''
  Nucl.\ Phys.\ B {\bf 266}, 709 (1986).
  doi:10.1016/0550-3213(86)90193-8
%
%
\bibitem{vandeVen1991} 
  A.~E.~M.~van de Ven,
  ``Two loop quantum gravity,''
  Nucl.\ Phys.\ B {\bf 378}, 309 (1992).
  doi:10.1016/0550-3213(92)90011-Y
%
%
\bibitem{Riess:1998cb}
  A.~G.~Riess {\it et al.} [Supernova Search Team],
  ``Observational evidence from supernovae for an accelerating universe and a cosmological constant,''
  Astron.\ J.\  {\bf 116} (1998) 1009
  doi:10.1086/300499
  [astro-ph/9805201].
%
%
\bibitem{Perlmutter:1998np}
  S.~Perlmutter {\it et al.} [Supernova Cosmology Project Collaboration],
  ``Measurements of Omega and Lambda from 42 high redshift supernovae,''
  Astrophys.\ J.\  {\bf 517} (1999) 565
  doi:10.1086/307221
  [astro-ph/9812133].
%
%
\bibitem{ArmendarizPicon:1999rj}
  C.~Armendariz-Picon, T.~Damour and V.~F.~Mukhanov,
  ``k - inflation,''
  Phys.\ Lett.\ B {\bf 458} (1999) 209
  doi:10.1016/S0370-2693(99)00603-6
  [hep-th/9904075].
%
%
\bibitem{ArmendarizPicon:2000dh}
  C.~Armendariz-Picon, V.~F.~Mukhanov and P.~J.~Steinhardt,
  ``A Dynamical solution to the problem of a small cosmological constant and late time cosmic acceleration,''
  Phys.\ Rev.\ Lett.\  {\bf 85} (2000) 4438
  doi:10.1103/PhysRevLett.85.4438
  [astro-ph/0004134].
%
%
\bibitem{Maggiore:2014sia} 
  M.~Maggiore and M.~Mancarella,
  ``Nonlocal gravity and dark energy,''
  Phys.\ Rev.\ D {\bf 90}, no. 2, 023005 (2014)
  doi:10.1103/PhysRevD.90.023005
  [arXiv:1402.0448 [hep-th]].
%
%
\bibitem{Stelle:1976gc} 
  K.~S.~Stelle,
  ``Renormalization of Higher Derivative Quantum Gravity,''
  Phys.\ Rev.\ D {\bf 16}, 953 (1977).
  doi:10.1103/PhysRevD.16.953.
%
%
\bibitem{Salam:1978fd} 
  A.~Salam and J.~A.~Strathdee,
  ``Remarks on High-energy Stability and Renormalizability of Gravity Theory,''
  Phys.\ Rev.\ D {\bf 18}, 4480 (1978).
  doi:10.1103/PhysRevD.18.4480
%
%
\bibitem{Julve:1978xn} 
  J.~Julve and M.~Tonin,
  ``Quantum Gravity with Higher Derivative Terms,''
  Nuovo Cim.\ B {\bf 46}, 137 (1978).
  doi:10.1007/BF02748637
%
%
\bibitem{Narain:2011gs} 
  G.~Narain and R.~Anishetty,
  ``Short Distance Freedom of Quantum Gravity,''
  Phys.\ Lett.\ B {\bf 711}, 128 (2012)
  doi:10.1016/j.physletb.2012.03.070
  [arXiv:1109.3981 [hep-th]].
%
%
\bibitem{Narain:2012nf} 
  G.~Narain and R.~Anishetty,
  ``Unitary and Renormalizable Theory of Higher Derivative Gravity,''
  J.\ Phys.\ Conf.\ Ser.\  {\bf 405}, 012024 (2012)
  doi:10.1088/1742-6596/405/1/012024
  [arXiv:1210.0513 [hep-th]].
%
%
\bibitem{Narain:2017tvp}
  G.~Narain,
  ``Signs and Stability in Higher-Derivative Gravity,''
  Int.\ J.\ Mod.\ Phys.\ A {\bf 33} (2018) no.04,  1850031
  doi:10.1142/S0217751X18500318
  [arXiv:1704.05031 [hep-th]].
%
%
\bibitem{Narain:2016sgk}
  G.~Narain,
  ``Exorcising Ghosts in Induced Gravity,''
  Eur.\ Phys.\ J.\ C {\bf 77} (2017) no.10,  683
  doi:10.1140/epjc/s10052-017-5249-z
  [arXiv:1612.04930 [hep-th]].
%
%
\bibitem{Codello:2006in}
  A.~Codello and R.~Percacci,
  ``Fixed points of higher derivative gravity,''
  Phys.\ Rev.\ Lett.\  {\bf 97} (2006) 221301
  doi:10.1103/PhysRevLett.97.221301
  [hep-th/0607128].
%
%
\bibitem{Niedermaier:2009zz}
  M.~R.~Niedermaier,
  ``Gravitational Fixed Points from Perturbation Theory,''
  Phys.\ Rev.\ Lett.\  {\bf 103} (2009) 101303.
  doi:10.1103/PhysRevLett.103.101303
%
%
\bibitem{Benedetti:2009rx}
  D.~Benedetti, P.~F.~Machado and F.~Saueressig,
  ``Asymptotic safety in higher-derivative gravity,''
  Mod.\ Phys.\ Lett.\ A {\bf 24} (2009) 2233
  doi:10.1142/S0217732309031521
  [arXiv:0901.2984 [hep-th]].
%
%
\bibitem{Salvio:2014soa} 
  A.~Salvio and A.~Strumia,
  ``Agravity,''
  JHEP {\bf 1406}, 080 (2014)
  doi:10.1007/JHEP06(2014)080
  [arXiv:1403.4226 [hep-ph]].
%
%
\bibitem{Lovelock:1971yv}
D.~Lovelock,
``The Einstein tensor and its generalizations,''
J. Math. Phys. \textbf{12} (1971), 498-501
doi:10.1063/1.1665613
%
%
\bibitem{Lovelock:1972vz}
D.~Lovelock,
``The four-dimensionality of space and the einstein tensor,''
J. Math. Phys. \textbf{13} (1972), 874-876
doi:10.1063/1.1666069
%
%
\bibitem{Lanczos:1938sf}
C.~Lanczos,
``A Remarkable property of the Riemann-Christoffel tensor in four dimensions,''
Annals Math. \textbf{39} (1938), 842-850
doi:10.2307/1968467
%
%
\bibitem{Glavan:2019inb}
D.~Glavan and C.~Lin,
``Einstein-Gauss-Bonnet gravity in 4-dimensional space-time,''
Phys. Rev. Lett. \textbf{124} (2020) no.8, 081301
doi:10.1103/PhysRevLett.124.081301
[arXiv:1905.03601 [gr-qc]].
%
%
\bibitem{Mardones:1990qc}
A.~Mardones and J.~Zanelli,
``Lovelock-Cartan theory of gravity,''
Class. Quant. Grav. \textbf{8} (1991), 1545-1558
doi:10.1088/0264-9381/8/8/018
%
%
\bibitem{Torii:2008ru}
T.~Torii and H.~a.~Shinkai,
``N+1 formalism in Einstein-Gauss-Bonnet gravity,''
Phys. Rev. D \textbf{78} (2008), 084037
doi:10.1103/PhysRevD.78.084037
[arXiv:0810.1790 [gr-qc]].
%
%
\bibitem{Kumar:2020uyz}
A.~Kumar and R.~Kumar,
``Bardeen black holes in the novel $4D$ Einstein-Gauss-Bonnet gravity,''
[arXiv:2003.13104 [gr-qc]].
%
%
\bibitem{Fernandes:2020rpa}
P.~G.~Fernandes,
``Charged Black Holes in AdS Spaces in $4D$ Einstein Gauss-Bonnet Gravity,''
[arXiv:2003.05491 [gr-qc]].
%
%
\bibitem{Kumar:2020owy}
R.~Kumar and S.~G.~Ghosh,
``Rotating black holes in the novel $4D$ Einstein-Gauss-Bonnet gravity,''
[arXiv:2003.08927 [gr-qc]].
%
%
\bibitem{Doneva:2020ped}
D.~D.~Doneva and S.~S.~Yazadjiev,
``Relativistic stars in 4D Einstein-Gauss-Bonnet gravity,''
[arXiv:2003.10284 [gr-qc]].
%
%
\bibitem{Aragon:2020qdc}
A.~Aragón, R.~Bécar, P.~González and Y.~Vásquez,
``Perturbative and nonperturbative quasinormal modes of 4D Einstein-Gauss-Bonnet black holes,''
[arXiv:2004.05632 [gr-qc]].
%
%
\bibitem{Mishra:2020gce}
A.~K.~Mishra,
``Quasinormal modes and Strong Cosmic Censorship in the novel 4D Einstein-Gauss-Bonnet gravity,''
[arXiv:2004.01243 [gr-qc]].
%
%
\bibitem{Heydari-Fard:2020sib}
M.~Heydari-Fard, M.~Heydari-Fard and H.~Sepangi,
``Bending of light in novel 4$D$ Gauss-Bonnet-de Sitter black holes by Rindler-Ishak method,''
[arXiv:2004.02140 [gr-qc]].
%
%
\bibitem{Nojiri:2020tph}
  S.~Nojiri and S.~D.~Odintsov,
  ``Novel cosmological and black hole solutions in Einstein and higher-derivative gravity in two dimensions,''
  EPL {\bf 130} (2020) no.1,  10004
  doi:10.1209/0295-5075/130/10004
  [arXiv:2004.01404 [hep-th]].
  %
  %
\bibitem{Hennigar:2020fkv}
R.~A.~Hennigar, D.~Kubiznak, R.~B.~Mann and C.~Pollack,
``Lower-dimensional Gauss--Bonnet Gravity and BTZ Black Holes,''
[arXiv:2004.12995 [gr-qc]].
%
%
\bibitem{Ghosh:2020vpc}
S.~G.~Ghosh and S.~D.~Maharaj,
``Radiating black holes in the novel 4D Einstein-Gauss-Bonnet gravity,''
[arXiv:2003.09841 [gr-qc]].
%
%
\bibitem{Casalino:2020kbt}
A.~Casalino, A.~Colleaux, M.~Rinaldi and S.~Vicentini,
``Regularized Lovelock gravity,''
[arXiv:2003.07068 [gr-qc]].
%
%
\bibitem{Konoplya:2020qqh}
R.~Konoplya and A.~Zhidenko,
``Black holes in the four-dimensional Einstein-Lovelock gravity,''
Phys. Rev. D \textbf{101} (2020) no.8, 084038
doi:10.1103/PhysRevD.101.084038
[arXiv:2003.07788 [gr-qc]].
%
%
\bibitem{Zhang:2020qam}
C.~Y.~Zhang, P.~C.~Li and M.~Guo,
``Greybody factor and power spectra of the Hawking radiation in the novel $4D$ Einstein-Gauss-Bonnet de-Sitter gravity,''
[arXiv:2003.13068 [hep-th]].
%
%
\bibitem{Konoplya:2020cbv}
R.~A.~Konoplya and A.~F.~Zinhailo,
``Grey-body factors and Hawking radiation of black holes in $4D$ Einstein-Gauss-Bonnet gravity,''
[arXiv:2004.02248 [gr-qc]].
%
%
\bibitem{Wei:2020poh}
S.~W.~Wei and Y.~X.~Liu,
``Extended thermodynamics and microstructures of four-dimensional charged Gauss-Bonnet black hole in AdS space,''
[arXiv:2003.14275 [gr-qc]].
%
%
\bibitem{Wei:2020ght}
S.~W.~Wei and Y.~X.~Liu,
``Testing the nature of Gauss-Bonnet gravity by four-dimensional rotating black hole shadow,''
[arXiv:2003.07769 [gr-qc]].
%
%
\bibitem{Roy:2020dyy}
R.~Roy and S.~Chakrabarti,
``A study on black hole shadows in asymptotically de Sitter spacetimes,''
[arXiv:2003.14107 [gr-qc]].
%
%
\bibitem{Guo:2020zmf}
M.~Guo and P.~C.~Li,
``The innermost stable circular orbit and shadow in the novel $4D$ Einstein-Gauss-Bonnet gravity,''
[arXiv:2003.02523 [gr-qc]].
%
%
\bibitem{Zeng:2020dco}
X.~X.~Zeng, H.~Q.~Zhang and H.~Zhang,
``Shadows and photon spheres with spherical accretions in the four-dimensional Gauss-Bonnet black hole,''
[arXiv:2004.12074 [gr-qc]].
%
%
\bibitem{Lu:2020iav}
H.~Lu and Y.~Pang,
``Horndeski Gravity as $D\rightarrow4$ Limit of Gauss-Bonnet,''
[arXiv:2003.11552 [gr-qc]].
%
%
\bibitem{Kobayashi:2020wqy}
T.~Kobayashi,
``Effective scalar-tensor description of regularized Lovelock gravity in four dimensions,''
[arXiv:2003.12771 [gr-qc]].
%
%
\bibitem{Mahapatra:2020rds}
S.~Mahapatra,
``A note on the total action of $4D$ Gauss-Bonnet theory,''
[arXiv:2004.09214 [gr-qc]].
%
%
\bibitem{Hennigar:2020lsl}
R.~A.~Hennigar, D.~Kubiznak, R.~B.~Mann and C.~Pollack,
``On Taking the $D\to 4$ limit of Gauss-Bonnet Gravity: Theory and Solutions,''
[arXiv:2004.09472 [gr-qc]].
%
%
\bibitem{1794944}
K.~Aoki, M.~A.~Gorji and S.~Mukohyama,
``A consistent theory of $D\rightarrow 4$ Einstein-Gauss-Bonnet gravity,''
[arXiv:2005.03859 [gr-qc]].
%
%
\bibitem{Deruelle:1989fj}
N.~Deruelle and L.~Farina-Busto,
``The Lovelock Gravitational Field Equations in Cosmology,''
Phys. Rev. D \textbf{41} (1990), 3696
doi:10.1103/PhysRevD.41.3696
%
%
\bibitem{Tangherlini:1963bw}
F.~Tangherlini,
``Schwarzschild field in n dimensions and the dimensionality of space problem,''
Nuovo Cim. \textbf{27} (1963), 636-651
doi:10.1007/BF02784569
%
%
\bibitem{Tangherlini:1986bw}
F.~Tangherlini,
``Dimensionality of Space and the Pulsating Universe,''
Nuovo Cim. \textbf{91} (1986), 209-217
%
%
\bibitem{Cognola:2013fva}
G.~Cognola, R.~Myrzakulov, L.~Sebastiani and S.~Zerbini,
``Einstein gravity with Gauss-Bonnet entropic corrections,''
Phys. Rev. D \textbf{88} (2013) no.2, 024006
doi:10.1103/PhysRevD.88.024006
[arXiv:1304.1878 [gr-qc]].
%
%
\bibitem{Feldbrugge:2017kzv}
J.~Feldbrugge, J.~L.~Lehners and N.~Turok,
``Lorentzian Quantum Cosmology,''
Phys. Rev. D \textbf{95} (2017) no.10, 103508
doi:10.1103/PhysRevD.95.103508
[arXiv:1703.02076 [hep-th]].
%
%
\bibitem{Feldbrugge:2017fcc}
J.~Feldbrugge, J.~L.~Lehners and N.~Turok,
``No smooth beginning for spacetime,''
Phys. Rev. Lett. \textbf{119} (2017) no.17, 171301
doi:10.1103/PhysRevLett.119.171301
[arXiv:1705.00192 [hep-th]].
%
%
\bibitem{nextpub}
G.~Narain, H.~Q.~Zhang,
in preparation.


\end{thebibliography}
\end{document}